\documentclass[showpacs,twocolumn,prd,aps]{revtex4}
\usepackage{amsfonts}
\usepackage{amsmath}
\usepackage{amsbsy}
\usepackage{amssymb}
\input{epsf}

\begin{document}

\title{Collision of strong gravitational and electromagnetic\\ waves in the expanding universe}

\author{G.A.~Alekseev}
     \email{G.A.Alekseev@mi.ras.ru}
\affiliation{\centerline{
\hbox{Steklov Mathematical
Institute of the Russian Academy of Sciences,}}\\
\centerline{\hbox{Gubkina str. 8, 119991, Moscow, Russia}}
}

\begin{abstract}
\noindent
An exact analytical model of the process of collision and nonlinear interaction of gravitational and/or electromagnetic soliton wave and strong non-soliton electromagnetic traveling wave of arbitrary profile propagating in the expanding universe (the symmetric Kasner spacetime) is presented.
In contrast to intuitive expectations that rather strong traveling waves can destroy the soliton, it occurs that the soliton survives during its interaction with electromagnetic wave of arbitrary amplitude and profile, but its parameters begin to evolve under the influence of this interaction. If a traveling electromagnetic wave possesses a finite duration, the soliton
parameters after interaction take constant values again, but these values in general are different from those before the interaction. Based on exact solutions of Einstein - Maxwell equations, our model demonstrates a series of nonlinear phenomena, such as (a) creation of gravitational waves in the collision of two electromagnetic waves, (b) creation of electromagnetic soliton wave in the collision of gravitational soliton with traveling
electromagnetic wave, (c) scattering of a part of soliton wave in the direction of propagation of traveling electromagnetic wave, (d) quasiperiodic oscillating character of fields in the wave interaction region and multiple mutual transformations of gravitational and electromagnetic waves in this region. The figures illustrate these features of nonlinear wave interactions in General Relativity.
\end{abstract}


\maketitle

\section{Introduction and results}

Since a beautiful discovery just a hundred years ago of Einstein's General Relativity theory, one of the most interesting problems in this theory is an understanding of different aspects of behaviour of strong gravitational fields and their interactions with various matter fields. The most part of such information known today was obtained using nonlinear geometric analysis of spacetime structures, approximation methods or numerical simulations \cite{Ashtekar:2013}. In most cases, however, this information possess too global and very restricted character with a lack of most interesting details.
On the other hand, the most detail information concerning properties of strong gravitational fields can be obtained from the exact solutions, especially if these include some constant or functional parameters providing a flexibility of these models.

During the centennial history of General Relativity a lot of exact solutions of Einstein equations were found, but there were no any solutions describing the  nonlinear interactions, mutual scattering and mutual transformations of arbitrary strong gravitational and electromagnetic waves propagating and colliding in curved spacetimes. All these phenomena are described by a large class of exact solutions of Einstein - Maxwell equations presented in this paper. Typical pictures of such interactions are shown on Fig.1 and Fig.2.

\begin{figure}[!]
\begin{center}
\epsfxsize=0.42\textwidth\epsfysize=0.22\textwidth \epsfbox{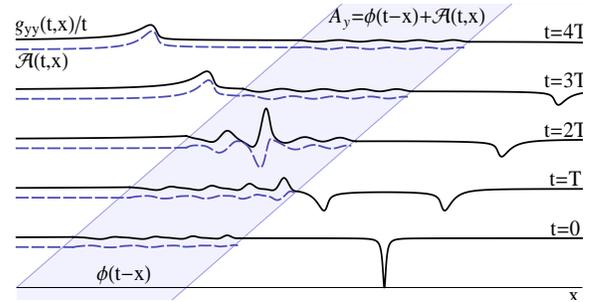}
\end{center}
\caption{\footnotesize Nonlinear interaction of \emph{gravitational soliton waves} (solid lines to the right from dark stripe) with traveling \emph{pure electromagnetic waves of arbitrary profile} $\phi(t-x)$ (dark stripe) in the symmetric Kasner spacetime background.
This interaction   creates gravitational (solid lines) and electromagnetic (dashed lines) waves in the waves interaction region (within the stripe). Oscillations of fields in this region, such that each local maximum of amplitude of one field is located near a local minimum of the other, mean the presence of multiple mutual transformations of gravitational and electromagnetic waves. After interaction (i.e. to the left from the stripe), gravitational part of soliton was changed and its electromagnetic counterpart was created. A part of soliton gravitational wave is scattered to positive x-direction (along the stripe).
}
\vspace{-5ex}
\end{figure}

First of all, we have to clarify that though in our class of exact solutions  the function $\phi(t-x)$ can be chosen arbitrarily, Fig.1 and Fig. 2 show the dynamics of waves for the simplest choice of this function for which $\phi(u)$ is a linear function in some interval $u_1<u<u_2$ and it possess constant values outside it. For other choices of $\phi(u)$ in this interval, the dynamics of fields is qualitatively similar to that shown on Fig.1 and Fig.2.

\vspace{0ex}
We have to notice also that "pure electromagnetic  waves" here and below mean
that in these waves the transverse components of metric ($g_{yy}$, $g_{yz}$, $g_{zz}$) are the same as for the background, i.e. the gravitational wave is absent. However, these electromagnetic waves produce some running longitudinal deformations of spacetime (of metric components on $(t,x)$-plane).

\begin{figure}[h]
\begin{center}
\epsfxsize=0.4\textwidth \epsfysize=0.28\textwidth\epsfbox{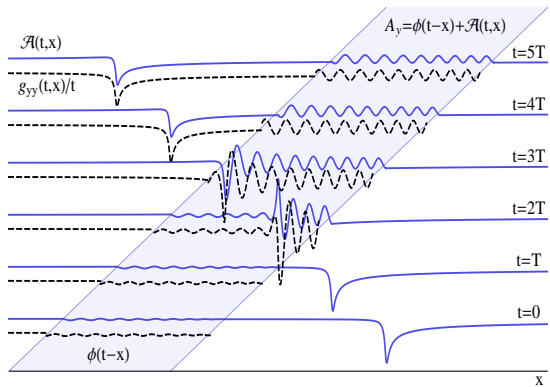}
\end{center}
\caption{\footnotesize Nonlinear interaction of \emph{electromagnetic soliton waves} (solid lines to the right from the dark stripe) with traveling  electromagnetic non-soliton waves of arbitrary profile $\phi(t-x)$ (within the stripe) in the symmetric Kasner spacetime background.
This interaction creates gravitational (dashed lines) and
electromagnetic (solid lines) waves in the waves interaction region (within the stripe). Oscillations of fields in this region, such that each local maximum of amplitude of one field is located near a local minimum of amplitude of the other, mean the presence of effect of multiple mutual
transformations of gravitational and electromagnetic waves. After interaction
(the region to the left from the stripe),  it occurs that the electromagnetic part of the soliton was changed and its gravitational counterpart was created.
A part of soliton wave was scattered to positive
x-direction (along the stripe).
}
\end{figure}

The Fig.1 and Fig. 2 show only the dynamics of one of the transverse components of metric $g_{yy}(t,x)$ and the component $A_y(t,x)$ of electromagnetic potential. The dynamics of other transverse metric components $g_{yz}$, $g_{zz}$ and of electromagnetic potential $A_z(t,x)$ are qualitatively similar to those shown above. The cases shown on the pictures are the subcases of our general class of solutions in which  the soliton may possess initially both gravitational and electromagnetic counterparts and the non-soliton electromagnetic wave may have any profile determined by an arbitrary function $\phi(t-x)$.

\subsection{On previous studies of interaction of gravitational and electromagnetic fields}

In the framework of perturbation theory, the first evidence to possibility of transformation of photons to gravitons in strong magnetic field was found by Gertsenshtein \cite{Gertsenshtein:1961}. Later, more detail information of this kind was obtained from the studies of a coupled system of small perturbations of gravitational and electromagnetic fields on charged black hole backgrounds. However, a discussion of a large literature of this kind and the corresponding specific results are not in the scope of the present paper. We mention here only the paper \cite{Alekseev-Sibgatullin:1974} where in WKB approximation it was shown that the interaction of gravitational and electromagnetic waves propagating in the external fields of charged black holes possesses a character of quasi-periodic mutual transformations, and the length of spatial modulation of each of these waves caused by such transformations was calculated. General perturbative approaches to studies of interaction of these waves in arbitrary external gravitational and electromagnetic fields in WKB approximation were suggested in \cite{Sibgatullin:1974} in coordinate form and in \cite{Alekseev-Khlebnikov:1978} in Newman - Penrose formalism.

The nonlinear interactions of gravitational and electromagnetic fields in different situations are described by a large variety of the known today exact solutions of Einstein-Maxwell equations. The most part of these solutions derived before the end of 70th had been collected in \cite{SKMHH:2003}. More detail discussions of physical and geometrical properties of these as well as of many later published solutions  were presented in \cite{Griffiths-Podolsky:2009}. Among the earliest examples are the exact solutions constructed in  \cite{Brill:1964, Batakis-Cohen:1972} for closed homogeneous anisotropic cosmological models which contain electromagnetic and scalar fields.

Many solutions for colliding plane waves in General Relativity which include the solutions for gravitational and electromagnetic waves, can be found in \cite{Griffiths:1991}. However, a great majority of these  solutions describe plane-fronted  waves with parallel rays colliding on the Minkowski background, where the mutual focusing of colliding waves and inevitable emergence of curvature singularities in the wave interaction region does not allow one to have a complete picture of wave dynamics in the whole spacetime until the null infinity. In contrast to these, in our solutions the background spacetime expansion prevents a creation of such singularities and provides a complete regular pictures of interacting wave dynamics.

Interesting class of solutions for gravitational and electromagnetic (and scalar) waves in closed cosmological models was constructed and discussed in \cite{Charach:1979, Charach-Malin:1980} as a subclass of more general class of solutions called as electromagnetic Einstein-Rosen waves and studied previously in \cite{Misra-Radhakrishna:1962, Thorne:1965, Melvin:1965}. In this class of solutions of Einstein - Maxwell equations, the field variables responsible for the transverse part of metric and for non-vanishing components of electromagnetic potential are considered as functionally dependent. The dynamical equations for these fields reduce to only one linear equation identical to the equation for pure vacuum Einstein - Rosen waves. In \cite{Charach:1979, Charach-Malin:1980} these solutions were used successfully for construction of standing waves in closed cosmological models \cite{footnote16}.

The class of electromagnetic Einstein-Rosen waves, unlike the class of solutions constructed in the present paper, can not provide us with a complete picture of the processes of collision of gravitational and electromagnetic waves. From physical point of view the solutions of this class do not carry some important imprints of the character of nonlinear interaction of gravitational and electromagnetic waves in general. The linearity of the dynamical equation for electromagnetic Einstein-Rosen waves means, in particular, that in the case of collision of incident waves from this class, the wave which leaved the interaction region do not carry any information about another wave which this wave had interacted with. The interaction of waves in this case emerges only in their influence on the longitudinal part of metric, but direct interactions of transverse modes and their mutual scattering are absent. Besides that, in the class of electromagnetic Einstein-Rosen waves, we can not consider such physically important situations as collision of two pure electromagnetic waves (i.e. of electromagnetic waves which are not accompanied by some transverse gravitational waves) or collision of pure gravitational and pure electromagnetic waves because in the electromagnetic Einstein-Rosen solutions the gravitational and electromagnetic parts of waves are in a fixed functional dependence of each other. However, all these difficulties are absent in the class of solutions constructed in the present paper.

The decades after the end of 70th were marked by discoveries of powerful methods for solution of Einstein's field equations such as soliton generating transformations and various integral equation methods (see \cite{Belinski-Verdaguer:2001, Alekseev:2011} and the references there).
However, the power of these methods (based on the integrability of Einstein's field equations for spacetimes with certain symmetries) have not been used in a full measure for construction of  non-trivial physically interesting solutions for interacting waves.
A rich new class of such solutions presented in this paper demonstrates the hidden yet power of these methods and allows us to elucidate a number of new interesting features of nonlinear dynamics of interacting waves.

\subsection{The main results}
In this paper, we present a class of exact solutions of Einstein-Maxwell equations  which describe the collision and nonlinear interaction of  one-soliton gravitational and electromagnetic waves and non-soliton pure electromagnetic traveling waves of arbitrary amplitudes and profiles propagating in the symmetric Kasner background. These solutions include a number of soliton parameters and an arbitrary real function of a null coordinate $u=t-x$
\[\ell+i s,\,\, c,\,\, d\quad\text{and}\quad \phi(u),
\]
which govern the initial location of a soliton along $x$-axis ($\ell$), its shape ($s$) and the amplitudes of its gravitational ($c$) and electromagnetic ($d$) components, while the arbitrarily chosen function $\phi(u)$ determines the amplitude and profile of  electromagnetic non-soliton wave. In physically most interesting cases, $\phi(u)$  takes constant values outside some interval $u_1<u<u_2$, i.e the non-soliton wave is of finite duration. In the simplest case, $\phi(u)$ is linear function of $u$ for $u_1<u<u_2$. Just this choice of $\phi(u)$ was used for non-soliton waves on Fig.1 and Fig.2.
\begin{itemize}
\item{These solutions demonstrate a qualitatively new picture of nonlinear interactions of soliton and non-soliton waves. In contrast to intuitive expectations that rather strong traveling non-soliton wave can destroy the soliton, it occurs that in the collision with non-soliton waves of arbitrary amplitudes and profiles, the soliton survives, but the values of its parameters $c$ and $d$  after the collision differ from the values of these parameters before this collision (compare the soliton wave profiles  to the left and to the right from the stripe on Fig.1 or on Fig.2).}
\item{Inside the wave interaction region $u_1<u<u_2$, the analytical form of the solution remains the same, but the soliton parameters $c$ and $d$, which were constants for $u<u_1$  begin to evolve: $c\to\widehat{c}(u)$, $d\to\widehat{d}(u)$ until the soliton leaves the wave interaction region, i.e. until  $u>u_2$, and so that
    \begin{equation}\label{chatdhat}
     \qquad\begin{array}{l}
    \widehat{c}= c\cosh S-i d\sinh S,\\[0.5ex]
    \widehat{d}= i c\sinh S+d\cosh S,
      \end{array}\hskip2ex S(u)=\sqrt{2}\displaystyle
\int\limits_{u_1}^{u}\dfrac{\phi^\prime(u)
d u}{\sqrt{w_o-u}}
    \end{equation}
    where $w_o=\ell+i s$. For $u< u_1$ where $\phi(u)=0$ we have $\widehat{c}=c$ and $\widehat{d}=d$. For $u>u_2$, where $\phi(u)=const$, these parameters become constants $\widehat{c}(u)=\widehat{c}(u_2)$, $\widehat{d}(u)=\widehat{d}(u_2)$ and the wave becomes pure soliton again. However, in general $\widehat{c}(u_2)\ne c$ and $\widehat{d}(u_2)\ne d$.}
\item{In the case $d=0$, the initial soliton ($u<u_1$) is pure vacuum. Then, for $u>u_2$ we have $\widehat{c}= c\cosh S(u_2)$ and $\widehat{d}= i c\sinh S(u_2)$. The condition $\widehat{d}\ne 0$ for $u>u_2$ means that some electromagnetic part of soliton wave was created in this interaction (see Fig.1).}
\item{In the case $c=0$, the initial soliton wave($u<u_1$) is pure electromagnetic. For $u>u_2$ we have $\widehat{c}= -i d\sinh S(u_2)$ and $\widehat{d}= d\cosh S(u_2)$. The condition $\widehat{c}\ne 0$ for $u>u_2$ means that a collision of soliton and non-soliton  electromagnetic waves created  gravitational part of soliton wave (see Fig.2).}
\item{In the wave interaction region $u_1<u<u_2$, the fields possess an oscillating character such that each local maximum of amplitude of incoming wave is located near a local minimum of amplitude of created  wave (see Fig.1 and Fig.2). These oscillations evince to the effect of multiple mutual transformations of gravitational and electromagnetic waves during their nonlinear interactions.}
\item{The nonlinear interaction of a soliton with non-soliton electromagnetic wave gives rise also to the effect of nonlinear scattering of waves so that a part of the initial soliton is pushed to propagate as gravitational and electromagnetic waves in the direction of propagation of the non-soliton wave.}
\end{itemize}

In the remaining part of the paper, we describe the symmetric Kasner background, the electrovacuum one-soliton waves and non-soliton traveling electromagnetic waves in this background as well as the new class of solutions for nonlinear interaction of these waves.

\section*{II NEW INTERACTING WAVE SOLUTIONS}
\setcounter{subsection}{0}
\subsection{The symmetric Kasner universe}
For the background where the waves will propagate and interact with its curvature and with each other, we choose a vacuum Kasner universe with two equal exponents $(p_1,p_2,p_3)=(-\frac 13,\frac 23,\frac23)$ which we take in the form
\begin{equation}\label{SKasner}
ds^2=-\displaystyle{\frac{1}{\sqrt{t}}}\,\, d u dv+t(dy^2+dz^2),
\end{equation}
where the null coordinates are $u=t-x$ and $v=t+x$.
For waves propagating in $x$-direction, the expansion with time of this background in the transverse directions  prevents a creation of caustics and curvature singularities and this allows to see a complete regular dynamics of waves, their  mutual scattering and transformations.

\subsection{Traveling non-soliton electromagnetic waves\\ in the symmetric Kasner universe}
The class of solutions of Einstein-Maxwell equations which describe  plane-fronted traveling non-soliton pure electromagnetic waves
with arbitrary amplitudes, profiles and polarizations
propagating in the symmetric Kasner universe along its axis of symmetry was constructed in \cite{Alekseev:2015}. In what follows, we consider only the case of such waves with linear polarizations. In this case
\begin{equation}\label{TravWaves}
\left\{\begin{array}{l}
ds^2=-\displaystyle{\frac{1}{\sqrt{t}}} e^{4 \mathcal{B}(u)}  du\, dv+t(dy^2+dz^2),\\[2ex]
A_i=\{0,\,0,\,\phi(u),\,0\},\quad \mathcal{B}(u)=\displaystyle\int {\phi\,{}^\prime}^2 (u)du
\end{array}\right.
\end{equation}
where $A_y=\phi(u)$ is an arbitrary real function of $u=t-x$.

\subsection{Soliton waves in the symmetric Kasner universe}
The waves of another type which may exist in the expanding Kasner universe (besides the traveling waves considered above)  are  the soliton gravitational and electromagnetic waves determined by soliton solutions of Einstein - Maxwell equations which were found in \cite{Alekseev:1980}. Omitting the details (and referring the readers for these to \cite{Alekseev:1980,Alekseev:1988}), we present here the electrovacuum one-soliton solution on the symmetric Kasner background (\ref{SKasner}):
\begin{equation}\label{KasnerSoliton}
\left\{\begin{array}{l}
ds^2=-f\, du\, dv+t \Bigl[h_{yy} dy^2+2 h_{yz}dy dz+h_{zz} dz^2\Bigr],\\[1ex]
A_i=\{0,\,0,\,A_y,\,A_z\},
\end{array}\right.
\end{equation}
where the metric functions possess the expressions
\begin{equation}\label{SolitonComponents}
\begin{array}{l}
h_{yy}=\dfrac{(\mathcal{G}-4 i s c) (\overline{\mathcal{G}}+ 4 i s\overline{c})}{\mathcal{D} \overline{\mathcal{D}}},\hskip3ex
h_{yz}=- \dfrac{4 s\,(c\,\overline{\mathcal{G}}+\overline{c}\,\mathcal{G})}{\mathcal{D} \overline{\mathcal{D}}},\\[2ex]
h_{zz}=\dfrac{(\mathcal{G}+4 i s c) (\overline{\mathcal{G}}-4 i s\overline{c})}{\mathcal{D} \overline{\mathcal{D}}},\hskip3ex
f=\dfrac 1{\sqrt{t}}\cdot\dfrac{\mathcal{D} \overline{\mathcal{D}}}{4\, r_+ r_- }
\end{array}
\end{equation}
and the electromagnetic potential components are the real parts of the corresponding complex potentials:
\begin{equation}\label{AyAz}
\{A_y,\,A_z\}=\Re\{\Phi,\widetilde{\Phi}\},\hskip1.5ex
\Phi=-2s d\,\dfrac{\mathcal{F}}{\mathcal{D}},\hskip1.5ex
\widetilde{\Phi}=2 i s d\,\dfrac{\widetilde{\mathcal{F}}}{\mathcal{D}}.
\end{equation}
In these expressions the following notations are used:
\begin{equation}\label{SolitonPolynoms}
\begin{array}{lcl}
\mathcal{D}=\overline{K}_+ K_--c\, \overline{c}\, K_+ \overline{K}_-+d\, \overline{d}\,\overline{K}_+\overline{K}_-,\\[1.2ex]
\mathcal{G}=K_+ K_--c\, \overline{c}\, \overline{K}_+ \overline{K}_-+d\, \overline{d}\,K_+\overline{K}_-,\\[1.2ex]
\mathcal{F}=\overline{K}_+-\overline{c} \overline{K}_-,\quad
\widetilde{\mathcal{F}}=\overline{K}_++\overline{c} \overline{K}_-.
\end{array}
\end{equation}
In these expressions, $K_\pm$ denote two complex functions
\begin{equation}\label{Kparameters}
K_+=\sqrt{R_+}+\dfrac{i s}{\sqrt{R_+}},\quad
K_-=\sqrt{R_-}+\dfrac{i s}{\sqrt{R_-}}
\end{equation}
and the real functions $R_\pm$ and  $r_\pm$ are defined as
\begin{equation}\label{RpRm}
\begin{array}{lcl}
R_+=\ell-u+r_+,&&r_+=\sqrt{(\ell-u)^2+s^2},\\[1ex]
R_-=\ell+v+r_-,&& r_-=\sqrt{(\ell+v)^2+s^2}.
\end{array}
\end{equation}
This solution depends on free constant parameters $\{\ell,\,s,\,c,\,d\}$
where the first two parameters are real and the other two are complex. The parameter $\ell$ determines a shift of the whole configuration along $x$-axis, $s$ determines the shape of the soliton, $c$ is responsible for gravitational and $d$ -- for electromagnetic parts of the soliton. In particular,
in the case $d=0$ we have pure vacuum one-soliton configuration which decays with time into two incident gravitational waves propagating in opposite $x$-directions (Fig.1, the region to the right from the stripe) and the case $c=0$ corresponds to pure electromagnetic one-soliton wave which propagates along the $x$-axis (Fig.2, the region to the right from the stripe).

\subsection{Interaction of gravitational and electromagnetic soliton waves with
electromagnetic\\ non-soliton traveling wave}
For construction of the solution, which describes the interaction of soliton waves (\ref{KasnerSoliton})--(\ref{RpRm}) with traveling waves (\ref{TravWaves}), we apply the electrovacuum soliton generating transformations \cite{Alekseev:1980,Alekseev:1988} choosing the spacetime  (\ref{TravWaves}) as the background for solitons. It is remarkable that the corresponding solution occurs to have the same form  (\ref{KasnerSoliton})--(\ref{RpRm}), but in all these expressions the soliton parameters $c$ and $d$ should be substituted by the evolving parameters $\widehat{c}(u)$ and $\widehat{d}(u)$ and, \emph{after this substitution}, the following corrections should be made for complex electromagnetic potentials in (\ref{AyAz}) and the conformal factor $f$ in (\ref{SolitonComponents}):
\[\begin{array}{lccl}
   c\to \widehat{c}(u),&&&\Phi\to\Phi+\phi(u),\\[0.5ex]
   d\to\widehat{d}(u),&&&\widetilde{\Phi}\to\widetilde{\Phi}+i\phi(u),
      \end{array}\quad f\to e^{4 \mathcal{B}(u)} f
\]
where $\widehat{c}(u)$, $\widehat{d}(u)$ and $\mathcal{B}(u)$ were defined in (\ref{chatdhat}) and (\ref{TravWaves}).

\section{Concluding remarks}
In this paper we present a new large class of exact solutions of Einstein - Maxwell equations which describe the nonlinear interaction of gravitational and electromagnetic soliton waves with non-soliton electromagnetic waves of arbitrary amplitudes and profiles propagating in (and interacting with) the  expanding spatially homogeneous universe represented by axisymmetric vacuum Kasner spacetime. It is worth to emphasize that the main purpose of this paper was not a construction of some viable cosmological model but the investigation of the process of collision and nonlinear interaction of arbitrarily strong gravitational and electromagnetic waves, their mutual scattering and mutual transformations.

From mathematical point of view, the choice of the expanding spacetime as the background for waves (instead of, e.g., the Minkowski spacetime) was very important because the expansion of this background prevents the emergence of caustics and curvature singularities caused by mutual focussing of colliding waves. This leads to a regular global picture of wave dynamics. Besides that, the choice of the axisymmetric Kasner background provides us with a rich class of solutions for traveling electromagnetic non-soliton waves which exist on this background. The solutions of this class are very simple and depend explicitly on an arbitrary function of a null coordinate which determines the arbitarily choosing amplitudes and profiles of these waves. (So large classes of explicitly calculated solutions for waves on the other backgrounds are not known.)

From physical point of view, however, the process of collision of such waves is a local process during which the global structure of the universe is not so important. Therefore, one may expect that the main features of this process will be similar for collision of high energy waves (with not too long durations) on the other backgrounds.

\section*{ACKNOWLEDGEMENTS}
This work is supported by the Russian Science Foundation under Grant No. 14-50-00005.


\begin{thebibliography}{99}
\bibitem{footnote16}{The author is thankful to anonymous referee for references [8,9] and [11,12] as well as for the suggestion to clarify the relation of the results of these papers to the results of the present paper.}
\end{thebibliography}
\end{document}